\documentclass[12pt]{iopart}

\usepackage{subeqn}
\usepackage{amssymb}

\def\QK{q^{\mbox{\tiny K}}}
\def\QC{q^{\mbox{\tiny C}}}
\def\QD{q}
\def\QG{q^{\mbox{\tiny G}}}
\def\QP{q}

\def\PK{p^{\mbox{\tiny K}}}

\def\PHK{\phi^{\mbox{\tiny K}}}
\def\PHC{\phi^{\mbox{\tiny C}}}
\def\PHD{\phi}
\def\PHG{\phi^{\mbox{\tiny G}}}

\def\CHK{\chi^{\mbox{\tiny K}}}
\def\CHC{\chi^{\mbox{\tiny C}}}
\def\CHD{\chi}
\def\CHG{\chi^{\mbox{\tiny G}}}

\def\QRK{q^{\mbox{\tiny K}\; \ast}}
\def\QHK{q^{\mbox{\tiny K}\; \dagger}}
\def\QRC{q^{\mbox{\tiny C}\; \ast}}
\def\QHC{q^{\mbox{\tiny C}\; \dagger}}
\def\QRG{q^{\mbox{\tiny G}\; \ast}}
\def\QHG{q^{\mbox{\tiny G}\; \dagger}}
\def\QRD{q^{\; \ast}}

\def\PCK{\phi^{\mbox{\tiny K}\; \ast}}
\def\PMK{\phi^{\mbox{\tiny K}\; \dagger}}
\def\PCC{\phi^{\mbox{\tiny C}\; \ast}}
\def\PMC{\phi^{\mbox{\tiny C}\; \dagger}}
\def\PCG{\phi^{\mbox{\tiny G}\; \ast}}
\def\PMG{\phi^{\mbox{\tiny G}\; \dagger}}
\def\PCD{\phi^{\; \ast}}

\def\RK{r^{\mbox{\tiny K}}}
\def\RC{r^{\mbox{\tiny C}}}
\def\RD{r}
\def\RG{r^{\mbox{\tiny G}}}
\def\RP{r}

\def\LK{L^{\mbox{\tiny K}}}
\def\MK{M^{\mbox{\tiny K}}}
\def\LGK{{\cal L}^{\mbox{\tiny K}}}

\def\LC{L^{\mbox{\tiny C}}}
\def\MC{M^{\mbox{\tiny C}}}

\def\LG{L^{\mbox{\tiny G}}}
\def\MG{M^{\mbox{\tiny G}}}

\def\det{\mbox{det}}

\newcommand{\CMP}{{\em Commun. Math. Phys.} }

\newcommand{\LMP}{{\em Lett. Math. Phys.} }
\newcommand{\LNC}{{\em Lett. Nuovo Cimento\/} }

\newcommand{\PhysS}{{\em Phys. Scr.} }
\newcommand{\PTP}{{\em Prog. Theor. Phys.} }
\newcommand{\PTPS}{{\em Prog. Theor. Phys. Suppl.} }

\newcommand{\SPJ}{{\em Sov. Phys.--JETP} }
\newcommand{\SAM}{{\em Stud. Appl. Math.} }

\newcommand{\TMP}{{\em Theor. Math. Phys.} }

\newcommand{\vecvar}[1]{\mbox{\boldmath$#1$}}

\def\hf{\frac{1}{2}}
\def\beq{\begin{equation}} \def\eeq{\end{equation}}
\def\bseq{\begin{subequations}} \def\eseq{\end{subequations}}
\def\bea{\begin{eqnarray}} \def\eea{\end{eqnarray}}
\def\bsea{\begin{subeqnarray}} \def\esea{\end{subeqnarray}}

\let\ti=\tilde

\def\eql{\eqalign}

\let\nn=\nonumber
\def\beann{\begin{eqnarray*}} \def\eeann{\end{eqnarray*}}

\let\a=\alpha  \let\g=\gamma \let\de=\delta
 \let\z=\zeta  
  \let\la=\lambda 
    \let\s=\sigma

  \let\D=\Delta

\newcommand{\poisson}[2]{\{#1\hspace{2pt}
\mbox{\raisebox{0.5pt}{$\otimes$}}\hspace{-6.3pt}
\mbox{\raisebox{-3pt}{,}}\hspace{5pt}#2\}}

\def\0{\over } \def\1{\vec }     \def\2{{1\over2}} \def\4{{1\over4}}
\def\5{\bar }  \def\6{\partial } \def\7#1{{#1}\llap{/}}

\def\<{\langle } \def\>{\rangle }

\def\i{{\rm i}} \def\tr{\mbox{tr}}

\def\e{{\rm e}}

\begin{document}
\jl{1}
\eqnobysec

\title{Integrable discretizations of derivative 
nonlinear Schr\"{o}dinger equations}

\author{Takayuki Tsuchida
}

\address{Graduate School of Mathematical Sciences, University of Tokyo, \\
3-8-1 Komaba, Meguro-ku, Tokyo 153-8914, Japan
}
\vspace*{3mm}
\address{E-mail: tsuchida@poisson.ms.u-tokyo.ac.jp}

\begin{abstract}
We propose integrable discretizations of derivative nonlinear 
Schr\"{o}dinger (DNLS) equations such as the Kaup--Newell equation, the 
Chen--Lee--Liu equation and the Gerdjikov--Ivanov equation 
by constructing Lax pairs. 
The discrete DNLS 
systems admit the reduction of complex 
conjugation between two dependent variables 
and possess 
bi-Hamiltonian structure. 
%
Through transformations of variables and reductions, 
we obtain novel integrable discretizations 
of the nonlinear Schr\"{o}dinger (NLS), modified KdV (mKdV), 
mixed NLS, matrix NLS, matrix KdV, matrix mKdV, coupled NLS, 
coupled Hirota, coupled Sasa--Satsuma and Burgers equations. 
We also discuss integrable discretizations of the sine-Gordon 
equation, the massive Thirring model and their generalizations. 
%
\end{abstract}

\maketitle

\section{Introduction}
\setcounter{equation}{0}
\label{}

The inverse scattering method (ISM) was invented 
by Gardner {\it et al.}\ 
\cite{GGKM} 
more than thirty years ago. 
They 
expressed the KdV equation as 
the compatibility condition of an eigenvalue problem and 
time evolution of the eigenfunction, and solved the KdV equation 
through the inverse problem of scattering. 
A pair of operators which defines the eigenvalue problem and 
the time evolution is now called the Lax pair. 
Zakharov and Shabat~\cite{ZS1,ZS2} considered 
a generalization of the 
eigenvalue problem and 
solved the nonlinear Schr\"{o}dinger (NLS) equation 
\beq
\i u_t + u_{xx} - 2uvu = O \hspace{7mm}
\i v_t - v_{xx} + 2vuv = O
\label{NLS}
\eeq
under some conditions via the ISM. 
It is noteworthy 
that the NLS equation \eref{NLS} is 
integrable 
for matrix-valued variables $u$ and $v$ \cite{Konope,Fordy1,Linden2,TW1}. 
Throughout this paper, we use the symbol italic $O$ 
when dependent variables in the 
equation considered can take their values in matrices. 
Ablowitz {\it et al.}\ \cite{AKNS} 
formulated the Zakharov--Shabat method in a plain manner 
and constructed 
the hierarchy of 
the NLS equation \eref{NLS}. 
Up to now, 
a variety of 
nonlinear evolution equations 
in continuous space-time 
have been solved via the ISM 
based on modified versions of the Zakharov--Shabat 
eigenvalue problem 
(see, {\it e.g.}\ \cite{Manakov,YO,Kaup,Takht,Kuz,KN2,KN,WKI1,WKI2}). 

In the remarkable paper \cite{AL1}, Ablowitz and Ladik 
proposed a natural discretization 
of the Zakharov--Shabat eigenvalue problem and obtained 
a discrete NLS hierarchy. 
Their success suggests that 
discrete integrable hierarchies are obtained 
through natural discretizations of the modified 
Zakharov--Shabat eigenvalue problems. 
It is, however, quite difficult 
to find such discretizations directly. In 
a few successful studies \cite{Hirotas,Orf,Ishimori2}, 
the difficulty is avoided skillfully with the help of 
alternative approaches. 
Hence, 
wide applicability of 
Lax-pair formulations 
has not been established 
in the theory of discrete integrable systems. 

In this paper we discuss integrable discretizations of 
derivative NLS (DNLS) equations and related systems 
by elaborating a new formulation of Lax pairs. 
Among the DNLS equations, the following three representatives 
are well-known \cite{Clark,Clark2,Clark3}, 
i.e.\ the Kaup--Newell equation \cite{KN},
\beq
\i \QK_t + \QK_{xx} - \i (\QK \RK \QK)_x = O 
\hspace{7mm}
\i \RK_t - \RK_{xx} - \i (\RK \QK \RK)_x = O
\label{KNeq}
\eeq
the Chen--Lee--Liu equation \cite{CLL},
\beq
\i \QC_t + \QC_{xx} - \i \QC_x \RC \QC = O
\hspace{7mm}
\i \RC_t - \RC_{xx} - \i \RC \QC \RC_x = O
\label{CLLeq}
\eeq
and the Gerdjikov--Ivanov equation \cite{GI},
\beq
\eql{\i \QG_t + \QG_{xx} + \i \QG \RG_x \QG 
+ \frac{1}{2} \QG \RG \QG \RG \QG = O
\\
\i \RG_t - \RG_{xx} + \i \RG \QG_x \RG 
- \hf \RG \QG \RG \QG \RG = O.
}
\label{GIeq}
\eeq
%
It has been shown that \eref{KNeq}--\eref{GIeq} are 
integrable 
for matrix-valued 
variables 
\cite{Linden1,Olver2,TW3} (see \cite{Sak,SWo} for 
complete lists of integrable coupled DNLS equations). 
The DNLS equations \eref{KNeq}--\eref{GIeq} correspond 
to special cases 
($\de = -1/2$, $-1/4$, $0$, respectively) 
of the generalized DNLS equation \cite{Kund}
\beq
\eql{\i q_t + q_{xx} + \i (4\de + 1) q^2 r_x + 4\i \de q q_x r 
+ (\de + 1/2) (4\de + 1) q^3 r^2 = 0 
\\
\i r_t - r_{xx} + \i (4\de + 1) r^2 q_x + 4\i \de r r_x q 
- (\de + 1/2) (4 \de + 1) r^3 q^2 = 0. 
}
\label{Keq}
\eeq
%
The zero symbol $0$ indicates that the dependent variables 
are restricted to scalars. 
We generate \eref{Keq} from the Gerdjikov--Ivanov equation \eref{GIeq} 
via the transformation 
%
\beq
q = \QG \exp{\Bigl( -2\i\de \int^x \QG \RG \,{\rm d}x' \Bigr)}
\hspace{7mm}
r = \RG \exp{\Bigl(  2\i\de \int^x \QG \RG \, {\rm d}x'\Bigr) }.
\label{cgt}
\eeq
A comprehensive description of physical applications 
of the DNLS equations as well as their exact solutions 
can be found in \cite{Clark2,Clark3}. 
An integrable discretization of the Chen--Lee--Liu 
equation \eref{CLLeq} was proposed by Date, Jimbo and Miwa \cite{Date}. 
However, 
their scheme does not admit 
the reduction of complex conjugation between $\QC_n$ and $\RC_n$. 
Therefore it is of little practical use. 
In other studies on discrete DNLS equations \cite{NCQ,Bus,Ward}, 
%
equations of motion are not explicitly given. 
Practically, little is known about natural integrable 
discretizations of the DNLS equations. 
%

To solve 
this problem, we consider discrete analogues of 
relations between 
the DNLS hierarchies 
and a generalization of the NLS 
hierarchy \cite{KN,Clark,GIK,WS}. 
We extend the 
Lax-pair formulation 
of Ablowitz and Ladik 
to obtain a generalization of the discrete NLS (Ablowitz--Ladik) 
system.  
We find out that 
integrable discretizations of the DNLS equations 
\eref{KNeq}--\eref{GIeq} 
together with their higher symmetries 
are embedded in the generalized Ablowitz--Ladik system. 
With appropriate gauge transformations, we obtain standard 
forms of eigenvalue problems for the discrete DNLS systems. 
Using a discrete version of the transformation 
\eref{cgt}, we obtain a discrete generalized DNLS system. 
%
All the discrete DNLS systems but one exception 
admit the reduction of complex 
conjugation. 
Thus they inherit the crucial 
property of the continuous DNLS systems. 
Through simple transformations and reductions for 
the discrete DNLS systems, we obtain 
integrable discretizations of various systems, 
{\it e.g.}\ the mixed NLS, matrix NLS, matrix KdV, 
matrix modified KdV (matrix mKdV), 
coupled Sasa--Satsuma and Burgers equations. 
In this way, 
we can derive integrable discretizations of surprisingly many systems 
from one generalization of the Ablowitz--Ladik formulation. 
This exemplifies that some approaches based on a Lax-pair formulation 
are very fruitful in the study of discrete integrable systems. 

The sine-Gordon equation and its generalization 
are integrable via the ISM based on 
the Zakharov--Shabat eigenvalue problem \cite{AKNS2,Eich}. 
In contrast, 
Ablowitz and Ladik were unsuccessful in obtaining 
a local discretization of the sine-Gordon equation 
(cf (2.10) in \cite{AL1}). 
This does not 
indicate that 
their discretization of the Zakharov--Shabat eigenvalue problem is 
useless in studying discrete sine-Gordon equations. 
It is just because 
time evolution of the eigenfunction assumed in \cite{AL1} is 
too general and inappropriate. 
We prove 
that discrete sine-Gordon equations and their generalizations 
are associated with the Ablowitz--Ladik eigenvalue problem 
(see~\ref{appB} and~\ref{appC}). 
The massive Thirring model and its variant types are known to 
have the same eigenvalue problems 
as the DNLS equations have \cite{Kuz,KN2,KN,Linden1,TW3,GIK}. 
In particular, the massive Thirring model and 
the Chen--Lee--Liu equation \eref{CLLeq} have the 
eigenvalue problem in common. 
This indicates that the massive Thirring-type models are related to some 
generalization of the sine-Gordon equation. 
Combining these items of information, 
we derive 
integrable discretizations of the massive Thirring-type models 
from the eigenvalue problems for the discrete DNLS systems. 

The paper consists of the following. In section 2, we 
show that the DNLS equations are embedded in a generalization 
of the NLS equation, and 
propose a generalization of the 
Ablowitz--Ladik system. 
In section 3, 
we reveal that 
space discretizations (for short, semi-discretizations) 
of the DNLS systems 
are embedded in the generalized Ablowitz--Ladik system. 
In section 4, we obtain a variety of 
integrable 
lattice systems related to the semi-discrete DNLS systems. 
In section 5, we investigate semi-discretizations of the 
massive Thirring-type models. 
The last section, section 6, is devoted to concluding remarks. 

In the present version of this paper, we do not discuss 
time discretizations of the semi-discrete DNLS systems. 
Some of the results on this problem can be 
found in old versions of 
this paper ({\tt arXiv:nlin.SI/0105053} ver.\ 1 or 2).

\section{Preliminaries}
\setcounter{equation}{0}
\label{}

In this section, 
we first show that the 
DNLS equations \eref{KNeq}--\eref{GIeq} are embedded in a 
generalization of the NLS equation \eref{NLS}. 
Next we 
extend the Ablowitz--Ladik formulation to obtain a semi-discretization 
of the generalized NLS system. 

\subsection{Embeddings of the DNLS equations into a generalized NLS equation}
\label{appA}

Let us begin with a brief description of Lax pairs.  
We consider a pair of linear equations 
\beq
\Psi_x = U \Psi$ \hspace{7mm}  $\Psi_t = V \Psi
\label{line0}
\eeq
for a column vector $\Psi$. The subscripts $x$ and $t$ denote the 
partial differentiation by $x$ and $t$ respectively. 
$U$ and $V$ are square matrices which depend on a parameter. 
The compatibility condition of \eref{line0} is given by 
\beq
U_t -V_x + UV-VU =O
\label{Lax_eq0}
\eeq
which we 
call the zero-curvature condition. If we choose the 
matrices $U$, $V$ appropriately, 
\eref{Lax_eq0} gives 
a system independent of the parameter. 
In such cases, the pair of matrices $U$, $V$ and 
the parameter are called the Lax pair and 
the spectral parameter, respectively. 

We introduce the following form of the Lax pair:
\bseq
\beq
\fl
U = \i \la \left[
\begin{array}{cc}
 -I_1 &  \\
  & I_2 \\
\end{array}
\right] +
\left[
\begin{array}{cc}
 w & u \\
 v & s \\
\end{array}
\right]
\label{}
\eeq
\beq
\fl
V = \i \la^2 \left[
\begin{array}{cc}
-2 I_1 & \\
   & 2 I_2 \\
\end{array}
\right]
+ \la \left[
\begin{array}{cc}
 & 2 u \\
2 v & \\
\end{array}
\right]
+\i \left[
\begin{array}{cc}
 c & u_x -wu + us \\
- v_x - vw + sv & d \\
\end{array}
\right].
\label{}
\eeq
\label{UVN}
\eseq
Here $\la$ is the spectral parameter. $U$ and $V$ are divided into 
four blocks as $(l_1+l_2) \times (l_1+l_2)$ matrices. 
$I_1$ and $I_2$ are, respectively, 
the $l_1 \times l_1$ and $l_2 \times l_2$ unit matrices. 
$w$, $s$, $u$, $v$ are matrix-valued variables. 
$c$ and $d$ are arbitrary functions 
at this stage. 
Substituting \eref{UVN} into the zero-curvature 
condition (\ref{Lax_eq0}), 
we obtain the following system (see (26) in \cite{Konop} 
for the case of scalar variables): 
\beq
\fl
\eql{\i w_t + c_x - wc + cw + (uv)_x + uvw - wuv = O
\\
\i s_t + d_x - sd + ds - (vu)_x - vus + svu = O
\\
\i u_t + u_{xx} -ud + cu - w_x u 
-2wu_x + 2u_x s + u s_x + w^2 u -2wus + u s^2 =O
\\
\i v_t - v_{xx} -vc + dv + s_x v 
+ 2 s v_x - 2v_x w - v w_x - s^2 v + 2svw - v w^2 =O. 
}
\label{gNLS}
\eeq
We call \eref{gNLS} the generalized NLS equation. 
If we set $w=O$, $s=O$, $c=-uv$, $d= vu$, 
\eref{gNLS} is reduced to the matrix NLS equation \eref{NLS}. 

We already know Lax pairs for the matrix DNLS equations 
\eref{KNeq}--\eref{GIeq} \cite{TW3}. 
With the help of gauge transformations, we can transform 
the Lax pairs 
into the form \eref{UVN} (see, {\it e.g.}\ \cite{Tsuchida4}). 
Thus the 
DNLS equations 
are embedded in the 
generalized NLS equation \eref{gNLS}. 
We omit details and show the main results in the following. 
For simplicity of the embedding formulae, 
we change the scalings of 
variables in \eref{KNeq}--\eref{GIeq}. 
\begin{enumerate}
\item[(a)]
If $\QK$ and $\RK$ satisfy the Kaup--Newell equation, 
\beq
\i \QK_t + \QK_{xx} +2 (\QK \RK \QK)_x = O 
\hspace{7mm}
\i \RK_t - \RK_{xx} +2 (\RK \QK \RK)_x = O
\label{KNeq2}
\eeq
$w$, $s$, $u$, $v$ defined by 
\[
w = - \QK \RK
\hspace{7mm}
s = \RK \QK
\hspace{7mm}
u = \QK
\hspace{7mm}
v = \RK_x - \RK \QK \RK
\]
satisfy the generalized NLS equation \eref{gNLS} with 
\[
c = -\QK_x \RK -2 \QK \RK \QK \RK
\hspace{7mm}
d = \RK \QK_x + 2 \RK \QK \RK \QK.
\]
\item[(b)]
If $\QC$ and $\RC$ satisfy the Chen--Lee--Liu equation, 
\beq
\i \QC_t + \QC_{xx} + 2 \QC_x \RC \QC = O
\hspace{7mm}
\i \RC_t - \RC_{xx} + 2 \RC \QC \RC_x = O
\label{CLLeq2}
\eeq
$w$, $s$, $u$, $v$ defined by 
\[
w = O
\hspace{7mm}
s = \RC \QC
\hspace{7mm}
u = \QC
\hspace{7mm}
v = \RC_x 
\]
satisfy the generalized NLS equation \eref{gNLS} with 
\[
c = -\QC \RC_x 
\hspace{7mm} 
d = \RC \QC_x + \RC \QC \RC \QC.
\]
\item[(c)]
If $\QG$ and $\RG$ satisfy the Gerdjikov--Ivanov equation,
\beq
\eql{\i \QG_t + \QG_{xx} -2 \QG \RG_x \QG 
 -2 \QG \RG \QG \RG \QG = O
\\
\i \RG_t - \RG_{xx} -2 \RG \QG_x \RG 
 +2 \RG \QG \RG \QG \RG = O
}
\label{GIeq2}
\eeq
$u$ and $v$ defined by 
\[
u = \QG
\hspace{7mm}
v = \RG_x + \RG \QG \RG
\]
satisfy the matrix NLS equation \eref{NLS}. 
%
\end{enumerate}
%

\subsection{Generalization of the Ablowitz--Ladik system}
\label{GeneAL}

We consider a semi-discrete version of the linear 
problem \eref{line0}:
\beq
\Psi_{n+1} = L_n \Psi_n \hspace{7mm} \Psi_{n,t} = M_n \Psi_n.
\label{line}
\eeq
The compatibility condition of (\ref{line}) is given by
\beq
 L_{n,t} +L_n M_n - M_{n+1}L_n = O
\label{Lax_eq}
\eeq
which is 
a semi-discrete version of the zero-curvature 
condition \eref{Lax_eq0}. 

We generalize the Lax pair 
proposed by Ablowitz and Ladik \cite{AL1} and consider 
the following form: 
\bea
\fl
  L_n = 
\left[
\begin{array}{cc}
z w_n  & u_n \\
 v_n  & \frac{1}{z} s_n \\
\end{array}
\right]
%
\hspace{7mm}
 M_n= 
\left[
\begin{array}{cc}
z^2 a I_1 + c_n  & za w_n^{-1} u_n + \frac{b}{z} u_{n-1} s_{n-1}^{-1} \\
za v_{n-1} w_{n-1}^{-1} + \frac{b}{z}s_n^{-1}v_n &  d_n + \frac{b}{z^2} I_2 \\
\end{array}
\right].
\label{AL_Lax}
\eea
%
Here $z$ is the spectral parameter and $a, b$ are constants. 
$L_n$ and $M_n$ are divided into four blocks 
as $(l_1+l_2) \times (l_1+l_2)$ matrices. 
$I_1$ and $I_2$ are 
unit matrices. 
$w_n$, $s_n$, $u_n$, $v_n$ are 
matrix-valued 
variables. 
$c_n$ and $d_n$ are arbitrary functions. 
Substitution of the Lax pair \eref{AL_Lax} 
into the zero-curvature condition \eref{Lax_eq} yields the following system:
\beq
\eql{w_{n,t} + w_n c_n - c_{n+1} w_n + a u_n v_{n-1} w_{n-1}^{-1} 
 - a w_{n+1}^{-1} u_{n+1} v_n = O
\\
s_{n,t} + s_n d_n - d_{n+1} s_n + b v_n u_{n-1} s_{n-1}^{-1} 
 - b s_{n+1}^{-1} v_{n+1} u_n = O
\\
u_{n,t} + u_n d_n - c_{n+1} u_n + b w_n u_{n-1} s_{n-1}^{-1} 
 - a w_{n+1}^{-1} u_{n+1} s_n = O
\\
v_{n,t} + v_n c_n - d_{n+1} v_n + a s_n v_{n-1} w_{n-1}^{-1} 
 - b s_{n+1}^{-1} v_{n+1} w_n = O.
}
\label{gAL}
\eeq
%
We call \eref{gAL} the generalized Ablowitz--Ladik system. 
When just two of $w_n$, $s_n$, $u_n$, $v_n$ are functionally 
independent, 
we may determine $c_n$, $d_n$ from the consistency of \eref{gAL}. 
If we set $w_n=I_1$, $s_n=I_2$, $c_n= -a (I_1 + u_n v_{n-1})$, 
$d_n= -b (I_2 + v_n u_{n-1})$, 
\eref{gAL} is reduced to the 
matrix Ablowitz--Ladik 
system \cite{GI2,Tsuchida2,Tsuchida3,Trubatch1,Vakh}:
%
\beq
\eql{u_{n,t} - a u_{n+1} + b u_{n-1} + (a-b) u_n 
 + a u_{n+1} v_n u_n 
 -b u_n v_n u_{n-1} 
= O
\\
v_{n,t} - b v_{n+1} + a v_{n-1} 
 + (b-a) v_n +b v_{n+1} u_n v_n 
 - a v_n u_n v_{n-1} 
= O.
}
\label{sd-matrix}
\eeq
%
The 
system \eref{sd-matrix} with $b=a^\ast$ admits the reduction of 
complex conjugation $v_n = \s u_{n}^\ast$ and not 
the reduction of Hermitian conjugation 
$v_n = \s u_{n}^\dagger$. 
Throughout this paper we use $\s$ to denote an arbitrary (but 
usually nonzero) real constant. 
We can construct an infinite set of conservation laws for \eref{gAL} 
by a recursive method \cite{Tsuchida2,Tsuchida3}. 
The first five conserved densities are 
\bea
\fl
\tr \log w_n \hspace{5mm} \tr \log s_n
\hspace{5mm} \tr \log (w_n - u_n s_n^{-1} v_n )
\hspace{5mm}
\tr (u_n v_{n-1} w_{n-1}^{-1} w_n^{-1}) 
\hspace{5mm}
\tr (v_n u_{n-1} s_{n-1}^{-1} s_n^{-1}).
\nn
\eea

The generalized Ablowitz--Ladik system \eref{gAL} 
gives a 
semi-discretization of the generalized NLS 
equation \eref{gNLS} 
together with its higher symmetry. 
To see this, we suppose the following scalings 
for \eref{gAL}:
\[
\fl
\eql{
w_n (t) = I_1 + \Delta w (x,t) \hspace{7mm} s_n (t) = I_2 + \D s (x,t) 
\hspace{7mm} u_n(t) = \D u (x,t) \hspace{7mm} v_n(t) = \D v (x,t) 
\\
a= -b= \frac{\i}{\D^2} \hspace{7mm} 
c_n (t)= -\frac{\i}{\D^2} I_1 + \i c(x,t) \hspace{7mm} 
d_n (t)= \frac{\i}{\D^2} I_2 + \i d(x,t). 
}
\]
Here $\D$ denotes the lattice spacing and $x = n \D$. 
In the continuum limit $\D \to 0$, \eref{gAL} 
is precisely reduced to \eref{gNLS}. 
A correspondence between the spectral parameters is 
given by $z = \exp (-\i \la \D)$. 
In the next section, 
we show that semi-discrete DNLS systems 
are embedded in \eref{gAL} through lattice analogues of the 
formulae (a)--(c) in section~\ref{appA}. 
%
We also mention that a system of semi-discrete 
coupled NLS equations is obtained 
as a reduction of \eref{gAL} \cite{Tsuchida3}. 

\section{Semi-discrete DNLS systems}
\setcounter{equation}{0}
\label{sdDNLS}

In this section, we obtain 
semi-discrete DNLS systems 
by considering lattice 
analogues of 
(a)--(c) for 
the generalized Ablowitz--Ladik system \eref{gAL}. 
Through a discrete analogue of the transformation \eref{cgt}, 
we obtain a semi-discrete 
generalized DNLS system. 

\subsection{Semi-discrete Kaup--Newell system}
\label{SubKN}

We find that a lattice analogue of 
(a) is given as follows.
\begin{enumerate}
\item[(A)]
If $\QK_n$ and $\RK_n$ satisfy the 
system
\bseq
\bea
&& 
\fl
\QK_{n,t} -a (I_1 - \QK_{n+1}\RK_{n+1})^{-1} \QK_{n+1}
+ a (I_1 - \QK_{n}\RK_{n})^{-1} \QK_{n}
- b (I_1 + \QK_{n}\RK_{n+1})^{-1} \QK_{n}
\nn \\
&& 
\mbox{}
+ b (I_1 + \QK_{n-1}\RK_{n})^{-1} \QK_{n-1} = O
\label{sdK1}
\\
&& \fl
\RK_{n,t} -b (I_2 + \RK_{n+1}\QK_{n})^{-1} \RK_{n+1}
+ b (I_2 + \RK_{n}\QK_{n-1})^{-1} \RK_{n}
- a (I_2 - \RK_{n}\QK_{n})^{-1} \RK_{n}
\nn \\
&& \mbox{}
+ a (I_2 - \RK_{n-1}\QK_{n-1})^{-1} \RK_{n-1} = O
\label{sdK2}
\eea
\label{sdKN}
\eseq
%
$w_n$, $s_n$, $u_n$, $v_n$ defined by 
\beq
\fl
w_n = I_1 - \QK_n \RK_n \hspace{7mm} s_n = I_2 +\RK_{n+1} \QK_n 
\hspace{7mm} u_n = \QK_n \hspace{7mm} 
v_n = \RK_{n+1} - \RK_n - \RK_{n+1} \QK_n \RK_n
\label{ws_KN}
\eeq
satisfy the generalized Ablowitz--Ladik system \eref{gAL} with 
\beq
\eql{c_n = -b I_1 -a (I_1 -\QK_n \RK_n )^{-1} + b (I_1 + \QK_{n-1} \RK_n )^{-1}
\\
d_n = -a I_2 +a (I_2 -\RK_n \QK_n )^{-1} - b (I_2 + \RK_{n} \QK_{n-1} )^{-1}.
}
\label{cd_KN}
\eeq
%
\end{enumerate}
The system \eref{sdKN} with $a=-b=\i$ gives 
an integrable semi-discretization of the Kaup--Newell 
equation \eref{KNeq2}, while 
\eref{sdKN} with $a=b=-1$ gives an integrable lattice version 
of a higher symmetry of \eref{KNeq2} (cf (3.39) in \cite{Linden1}). 

To consider the reduction of complex conjugation or 
Hermitian conjugation, we suppose that 
$\QK_{n}$ and $\RK_n$ 
are defined on 
the fractional lattice $n \in {\mathbb Z}/2$.
If $b=a^\ast$, the semi-discrete Kaup--Newell system 
\eref{sdKN} admits either the reduction 
$\RK_n = \i \s \QRK_{n-\hf}$ 
or the reduction 
$\RK_n = \i \s \QHK_{n-\hf}$. 
%
This property is crucial for various 
applications of 
\eref{sdKN}. 

Through the gauge transformation
\bea
\Phi_n = 
\left[
\begin{array}{cc}
f(z) I_1  &  O \\
 -\frac{1}{z} \RK_n  &  I_2 \\
\end{array}
\right] \Psi_n
\label{gaugeKN}
\eea
where $f(z)$ is an arbitrary function of $z$, 
the 
linear problem \eref{line} and 
the Lax pair \eref{AL_Lax} with \eref{ws_KN} and \eref{cd_KN} 
are changed into 
%
\[
\Phi_{n+1} = \LK_n \Phi_n \hspace{7mm}  
\Phi_{n,t} = \MK_n \Phi_n
\]
and 
%
\bseq
\bea
\fl
\LK_n = \left[
\begin{array}{cc}
z I_1 - ( z-\frac{1}{z}) \QK_n \RK_n 
	& f(z) \QK_n \\
 \frac{1}{f(z)} (-1 + \frac{1}{z^2} ) \RK_n  &  \frac{1}{z} I_2 \\
\end{array}
\right]
= 
\left[
\begin{array}{cc}
z I_1 & z f(z) \QK_n \\
 O &  I_2 \\
\end{array}
\right]
\left[
\begin{array}{cc}
 I_1 & O \\
 \frac{1}{f(z)} (-1 + \frac{1}{z^2} ) \RK_n  &  \frac{1}{z} I_2 \\
\end{array}
\right]
\nn \\
\label{Ln_KN}
\\
%
\fl
\MK_n = 
\left[
\begin{array}{c|c}
\begin{array}{l}
[a(z^2 -1) + b(-1+\frac{1}{z^2})] I_1  
\\
 \mbox{} + b(1-\frac{1}{z^2}) (I_1 + \QK_{n-1}\RK_n)^{-1} 
\end{array}
&
\begin{array}{l}
f(z) [az (I_1-\QK_n \RK_n)^{-1} \QK_n 
\\
 \mbox{} + \frac{b}{z} (I_1 + \QK_{n-1} \RK_n)^{-1} \QK_{n-1} ] 
\end{array}
\\
\hline 
\begin{array}{l}
\frac{1}{f(z)} [ a (-z + \frac{1}{z}) 
 (I_2- \RK_{n-1} \QK_{n-1})^{-1}\RK_{n-1} 
\\
 \mbox{}+ b(- \frac{1}{z} + \frac{1}{z^3}) 
(I_2 + \RK_n \QK_{n-1})^{-1} \RK_n ] 
\end{array}
&
b (-1 + \frac{1}{z^2}) (I_2 + \RK_n \QK_{n-1})^{-1}
\end{array}
\right].
\label{Mn_KN}
\eea
\label{Lax_KN}
\eseq
We can check 
that substitution of the 
Lax pair \eref{Lax_KN} into the zero-curvature 
condition \eref{Lax_eq} yields the semi-discrete 
Kaup--Newell system \eref{sdKN}. If we set 
\bea
\fl
f(z) = \bigl( -1 + \frac{1}{z^2} \bigr)^{\hf} \hspace{7mm}
z = \exp (-\i \z^2 \Delta) \hspace{7mm} 
\QK_n = \bigl( - \frac{\i \Delta}{2} \bigr)^{\hf} q_n \hspace{7mm}
\RK_n = \bigl( - \frac{\i \Delta}{2} \bigr)^{\hf} r_n 
\nn
\eea
%
the $L_n$-matrix \eref{Ln_KN} has the following asymptotic form:
\[
\LK_n = 
\left[
\begin{array}{cc}
I_1  &  \\
  & I_2 \\
\end{array}
\right] + 
\Delta \left[
\begin{array}{cc}
 -\i \z^2 I_1 & \z q_n \\
  \z r_n & \i \z^2 I_2 \\
\end{array}
\right]
+ O(\Delta^2). 
\]
Consequently, in the continuum limit $\Delta \to 0$, 
we recover the 
eigenvalue problem 
proposed by Kaup and Newell \cite{KN}. 

For the time being, we assume without loss of generality that 
$\QK_n$ and $\RK_n$ are square matrices. 
If necessary, we append dummy 
variables 
to $\QK_n$ and $\RK_n$ which are finally equalized 
to zero. 
Then a set of Hamiltonian and Poisson brackets 
for the semi-discrete 
Kaup--Newell system \eref{sdKN} is given by 
\bseq
\bea
H = \sum_n \bigl[ -a \log \det (I - \QK_n \RK_n) 
 + b \log \det (I + \QK_n \RK_{n+1})
\bigr]
\label{H_KN}
\eea
and
\bea
\poisson{\QK_n}{\QK_m} = 
\poisson{\RK_n}{\RK_m} = O \hspace{5mm}
\poisson{\QK_n}{\RK_m} = (\de_{n+1,m} - \de_{n,m})\Pi.
\eea
\label{firstH}
\eseq
Here $\Pi$ denotes the permutation matrix: 
$\Pi_{kl}^{ij} = \de_{i,l}\de_{j,k}$. 
It is easily proven that $\QK_{n,t} = \{ \QK_n, H \}$ and 
$\RK_{n,t} = \{ \RK_n, H \}$ coincide with \eref{sdK1} and 
\eref{sdK2}, respectively. 

Let us prove that in the periodic boundary case 
($\QK_{n+M} = \QK_n$, $\RK_{n+M} = \RK_n$) 
\eref{sdKN} has an involutive set of 
conserved quantities. 
We introduce 
new variables $\PK_n$ by 
$\QK_n = \PK_{n+1} - \PK_n$ and $\PK_{n+M} = \PK_n$, 
%
which satisfy ultra-local Poisson brackets: 
\bea
\poisson{\PK_n}{\PK_m} = 
\poisson{\RK_n}{\RK_m} = O \hspace{5mm}
\poisson{\PK_n}{\RK_m} = \de_{n,m}\Pi. 
\nn
\eea
For convenience, we set $f(z) = 1/z$. 
Applying a gauge transformation to the 
$L_n$-matrix \eref{Ln_KN}, we obtain a new 
ultra-local $L_n$-matrix:
%
\bea
\LGK_n (z) &=&
\left[
\begin{array}{cc}
- I  &  \PK_{n+1} \\
 O &  I \\
\end{array}
\right]
\LK_n 
\left[
\begin{array}{cc}
 -I &  \PK_{n} \\
  O &  I \\
\end{array}
\right]
\nn \\
&=& \left[
\begin{array}{cc}
zI  &  \\
  & \frac{1}{z} I\\ 
\end{array}
\right]
+ 
\bigl( z-\frac{1}{z} \bigr) 
\left[
\begin{array}{cc}
 \PK_n \RK_n & - \PK_n - \PK_n \RK_n \PK_n \\
 \RK_n  & -\RK_n \PK_n \\
\end{array}
\right].
\nn
\eea
The matrix $\LGK_n (z)$ satisfies the 
fundamental $r$-matrix relation: 
\beq
\poisson{\LGK_n (\la)}{\LGK_m (\mu)} 
= \de_{n,m} [\LGK_n (\la) \otimes \LGK_m (\mu), \, r(\la, \mu) ].
\label{fund}
\eeq
Here 
$r(\la, \mu)$ in the case of scalar variables 
is given by 
%
\beq
r(\la, \mu) = - \frac{(\la^2 -1)(\mu^2-1)}{\la^2 - \mu^2} 
\left[
\begin{array}{cccc}
 1 &  &  &  \\
  & 0 & 1 &  \\
  & 1 & 0 &  \\
  &  &  & 1 \\
\end{array}
\right].
\label{r_sca}
\eeq
%
$r(\la,\mu)$ for matrix-valued variables 
is given by replacing the ones in 
the matrix \eref{r_sca} 
with the permutation matrix $\Pi$. This is similar 
to the $r$-matrix given in \cite{TW1}. 

The fundamental $r$-matrix relation \eref{fund} results in 
commutativity of the traces of the monodromy matrix 
$\LGK_M \LGK_{M-1} \cdots \LGK_1$ for different values 
of the spectral parameter. 
This generates a set of 
$M$ conserved quantities in involution. 
Owing to the 
relation 
$\tr\, ( \LGK_M \LGK_{M-1} \cdots \LGK_1 )
= \tr\, ( \LK_M \LK_{M-1} \cdots \LK_1 ) $ we can write 
the conserved quantities 
in terms of 
the original variables $\QK_n$ and $\RK_n$. 
The set of conserved quantities includes 
the Hamiltonian \eref{H_KN} itself. 

\subsection{Semi-discrete Chen--Lee--Liu system}
\label{SubCLL}

We find that a lattice analogue of (b) is given 
as follows.
%
\begin{enumerate}
\item[(B)]
If $\QC_n$ and $\RC_n$ satisfy the system
\beq
\hspace{-8mm}
\eql{\QC_{n,t} - a(\QC_{n+1} - \QC_{n})(I_2 + \RC_{n} \QC_{n})
- b(\QC_{n} - \QC_{n-1}) (I_2 + \RC_{n} \QC_{n-1})^{-1} = O
\\
\RC_{n,t} -b (I_2 + \RC_{n+1} \QC_n)^{-1} (\RC_{n+1}-\RC_n) 
-a (I_2 + \RC_{n} \QC_{n})(\RC_{n}-\RC_{n-1}) =O
}
\label{sdCLL}
\eeq
$w_n$, $s_n$, $u_n$, $v_n$ defined by 
\beq
\hspace{-8mm}
w_n = I_1  \hspace{7mm}
s_n = I_2 + \RC_{n+1} \QC_{n} \hspace{7mm}
u_n = \QC_n  \hspace{7mm}
v_n = \RC_{n+1} - \RC_n
\label{ws_CLL}
\eeq
satisfy the generalized Ablowitz--Ladik system \eref{gAL} with 
\beq
\hspace{-8mm}
c_n = -a I_1 - a \QC_n (\RC_n - \RC_{n-1}) 
\hspace{7mm}
d_n = - b (I_2 + \RC_{n} \QC_{n-1} )^{-1} + a \RC_n \QC_n.
\label{cd_CLL}
\eeq
%
\end{enumerate}
The system \eref{sdCLL} with $a=-b=\i$ gives 
an integrable semi-discretization of the Chen--Lee--Liu 
equation \eref{CLLeq2}, while 
\eref{sdCLL} with $a=b=-1$ gives 
an integrable lattice version of a 
higher symmetry of \eref{CLLeq2} (cf (3.37) in \cite{Linden1}). 

Unfortunately, the semi-discrete Chen--Lee--Liu system 
\eref{sdCLL} admits neither the reduction 
$\RC_n = \i \s \QRC_{n-k}$ nor the reduction 
$\RC_n = \i \s \QHC_{n-k}$. 
This is an intrinsic drawback for practical use. 
We obtain an alternative semi-discretization of 
the Chen--Lee--Liu system, which admits the reduction 
of complex conjugation, in section~\ref{AlCLL}.

Through the gauge transformation \eref{gaugeKN} with 
replacing $\RK_n$ by $\RC_n$, 
%
the linear problem \eref{line} and 
the Lax pair \eref{AL_Lax} with \eref{ws_CLL} and 
\eref{cd_CLL} are changed into 
%
\[
\Phi_{n+1} = \LC_n \Phi_n \hspace{7mm}
\Phi_{n,t} = \MC_n \Phi_n
\]
and
\bseq
\beq
\fl
\LC_n = \left[
\begin{array}{cc}
z I_1 + \frac{1}{z} \QC_n \RC_n & f(z) \QC_n \\
 \frac{1}{f(z)} (-1 + \frac{1}{z^2} ) \RC_n  &  \frac{1}{z} I_2 \\
\end{array}
\right]
\label{Ln_CLL}
\eeq
\bea
\fl
\MC_n = 
\left[
\begin{array}{c|c}
\begin{array}{l}
[a(z^2 -2) + \frac{b}{z^2}] I_1 + a (I_1 + \QC_{n} \RC_{n-1})
\\
\mbox{} -\frac{b}{z^2} (I_1 + \QC_{n-1}\RC_n)^{-1} 
\end{array}
&
\begin{array}{l}
f(z) [az \QC_n 
\\
 \mbox{} 
+ \frac{b}{z} (I_1 + \QC_{n-1} \RC_n)^{-1} \QC_{n-1} ] 
\end{array}
\\
\hline 
\begin{array}{l}
\frac{1}{f(z)} [ a (-z + \frac{1}{z}) \RC_{n-1} 
\\
 \mbox{}+ b(- \frac{1}{z} + \frac{1}{z^3}) 
(I_2 + \RC_n \QC_{n-1})^{-1} \RC_n ] 
\end{array}
&
b (-1 + \frac{1}{z^2}) (I_2 + \RC_n \QC_{n-1})^{-1}
\end{array}
\right].
\label{Mn_CLL}
\eea
\label{Lax_CLL}
\eseq
We can prove that substitution of the 
Lax pair \eref{Lax_CLL} into the zero-curvature 
condition \eref{Lax_eq} yields the semi-discrete 
Chen--Lee--Liu system \eref{sdCLL}. 
With appropriate scalings, we 
reproduce the eigenvalue problem for 
the Chen--Lee--Liu system 
in the continuum limit. 
%
%

\subsection{Semi-discrete Gerdjikov--Ivanov system}
\label{SubGI}

We find that a lattice analogue of (c) is given 
as follows.
\begin{enumerate}
\item[(C)]
If $\QG_n$ and $\RG_n$ satisfy the system
\bseq
\bea
&& \fl 
 \QG_{n,t} - a \QG_{n+1} + b \QG_{n-1} + (a-b) \QG_n
 + a \QG_{n+1} (\RG_{n+1}-\RG_n) \QG_n 
- b \QG_{n} (\RG_{n+1}-\RG_n) \QG_{n-1} 
\nn \\
&&
\mbox{} 
 + a \QG_{n+1}\RG_{n+1}\QG_{n}\RG_{n}\QG_{n}
 - b \QG_{n}\RG_{n+1}\QG_{n}\RG_{n}
 \QG_{n-1} = O 
\\
&& \fl
\RG_{n,t} - b \RG_{n+1} + a \RG_{n-1} + (b-a) \RG_n
 - b \RG_{n+1} (\QG_{n}-\QG_{n-1}) \RG_n 
 + a \RG_{n} (\QG_{n}-\QG_{n-1}) \RG_{n-1} 
\nn \\ 
&& 
\mbox{} 
 + b \RG_{n+1}\QG_{n}\RG_{n} \QG_{n-1} \RG_{n}
 - a \RG_{n}\QG_{n}\RG_{n}\QG_{n-1}\RG_{n-1} = O
\eea
\label{sdGI}
\eseq
%
$u_n$ and $v_n$ defined by 
\beq
u_n = \QG_n  \hspace{7mm}
v_n = \RG_{n+1} - \RG_n + \RG_{n+1} \QG_n \RG_n
\label{ws_GI}
\eeq
satisfy the matrix Ablowitz--Ladik system \eref{sd-matrix}.
\end{enumerate}
If $\QG_n = I$, the relation 
\eref{ws_GI} between $\RG_n$ and $v_n$ 
is nothing but a discrete Miura transformation \cite{Wad}. 
%
%
The system \eref{sdGI} with $a=-b=\i$ gives 
an integrable semi-discretization of the Gerdjikov--Ivanov 
equation \eref{GIeq2}, while \eref{sdGI} with $a=b=-1$ 
gives an integrable lattice version of a 
higher symmetry of \eref{GIeq2} (cf (3.45) in \cite{Linden1}). 
The semi-discrete Gerdjikov--Ivanov system \eref{sdGI} 
with $b=a^\ast$ admits the 
reduction 
$\RG_n = \i \s \QRG_{n-\hf}$ 
and not the reduction $\RG_n = \i \s \QHG_{n-\hf}$. 

Through the gauge transformation \eref{gaugeKN} with 
replacing $\RK_n$ by $\RG_n$, 
%
the linear problem \eref{line} and 
the Lax pair 
for \eref{sd-matrix} 
are changed into 
%
\[
\Phi_{n+1} = \LG_n \Phi_n 
\hspace{7mm}  
\Phi_{n,t} = \MG_n \Phi_n
\]
and
\bseq
\beq
\fl
\LG_n = \left[
\begin{array}{cc}
z I_1 + \frac{1}{z} \QG_n \RG_n 
	& f(z) \QG_n \\
 \frac{1}{f(z)} (-1 + \frac{1}{z^2} ) 
	(\RG_n - \RG_{n+1} \QG_n \RG_n)
	&  \frac{1}{z} I_2 - \frac{1}{z} \RG_{n+1}\QG_n \\
\end{array}
\right]
\label{LG_n}
\eeq
\bea
\fl
\MG_n = 
\left[
\begin{array}{c|c}
\begin{array}{l}
 a(z^2 -1) I_1 + a\QG_n \RG_{n-1} 
\\
 \mbox{} 
+ \frac{b}{z^2} \QG_{n-1} \RG_n 
 - a \QG_n \RG_n \QG_{n-1} \RG_{n-1}
\end{array}
&
f(z) (az \QG_n + \frac{b}{z} \QG_{n-1})
\\
\hline 
\begin{array}{l}
\frac{1}{f(z)} (-1 + \frac{1}{z^2}) 
(I_2 - \RG_{n} \QG_{n-1})
\\
\mbox{} \times
(a z \RG_{n-1} + \frac{b}{z} \RG_n) 
\end{array}
&
\begin{array}{l}
b (-1 + \frac{1}{z^2}) (I_2 - \RG_n \QG_{n-1}) 
 - a \RG_n \QG_n 
\\
\mbox{}
- b \RG_{n+1} \QG_{n-1} -b \RG_{n+1} \QG_n \RG_n \QG_{n-1}
\end{array}
\end{array}
\right].
\label{Mn_GI}
\eea
\label{Lax_GI}
\eseq
%
Putting the Lax pair \eref{Lax_GI} into the zero-curvature 
condition \eref{Lax_eq}, we exactly obtain the 
semi-discrete Gerdjikov--Ivanov system \eref{sdGI}. 
The $L_n$-matrix \eref{LG_n} 
depends on $\QG_n$, $\RG_{n}$ and $\RG_{n+1}$, 
thus it is not ultra-local. 
It is an open question whether the system \eref{sdGI} 
possesses a Lax pair 
with an ultra-local $L_n$-matrix. 
In the continuum limit of space, 
we regenerate the eigenvalue problem for the 
Gerdjikov--Ivanov system. 
%

By a generalization of 
the transformation \eref{ws_GI}
\[
u_n = \QP_n
\hspace{7mm}
v_n = \la \RP_{n+1} - \RP_n + \la \RP_{n+1} \QP_n \RP_n
\]
for 
\eref{sd-matrix}, 
we obtain a generalization of 
\eref{sdGI}:
\bseq
\bea
&& \fl 
 \QP_{n,t} - a \QP_{n+1} + b \QP_{n-1} + (a-b) \QP_n
 + a \QP_{n+1} (\la \RP_{n+1} - \RP_n) \QP_n 
 - b \QP_{n} (\la \RP_{n+1} - \RP_n) \QP_{n-1} 
\nn \\
&&
\mbox{} 
 + a \la \QP_{n+1}\RP_{n+1}\QP_{n}\RP_{n}\QP_{n}
 - b \la \QP_{n}\RP_{n+1}\QP_{n}\RP_{n} \QP_{n-1} = O 
\\
&& \fl
\RP_{n,t} - b \RP_{n+1} + a \RP_{n-1} + (b-a) \RP_n
 - b \RP_{n+1} (\QP_{n}- \la \QP_{n-1}) \RP_n 
 + a \RP_{n} (\QP_{n}- \la \QP_{n-1}) \RP_{n-1} 
\nn \\ 
&& 
\mbox{} 
 + b \la \RP_{n+1}\QP_{n}\RP_{n} \QP_{n-1} \RP_{n}
 - a \la \RP_{n}\QP_{n}\RP_{n}\QP_{n-1}\RP_{n-1} = O.
\eea
\label{sdGI2}
\eseq
%
This system 
is considered as a one-parameter 
deformation of the matrix Ablowitz--Ladik system \eref{sd-matrix}. 
The change of variables 
$\QP_n = \QG_n \la^n \e^{[(\la-1) a + (1-\la^{-1}) b ]t}$, 
$\RP_n = \RG_n \la^{-n} \e^{-[(\la-1) a + (1-\la^{-1}) b ]t}$ 
and scalings of $a$, $b$ 
cast \eref{sdGI2} into 
\eref{sdGI}. 
Transformations of this kind are very useful 
for applications of the discrete DNLS systems 
and we utilize them in sections~\ref{sdmNLS}--\ref{SDNLS}.
%


\subsection{Semi-discrete generalized DNLS system}
\label{subG}
%
In sections~\ref{subG}--\ref{AlCLL}, 
we assume that dependent variables are scalar. 
As a discrete analogue of 
\eref{cgt}, 
we consider the following transformation for 
the semi-discrete Gerdjikov--Ivanov system \eref{sdGI}: 
%
\beq
q_n = \QG_n \prod_{j= -\infty}^n \Bigl( 
\frac{1 - \QG_{j-1}\RG_{j}}{1 + \QG_{j}\RG_{j}}
\Bigr)^{-2\gamma} 
\hspace{7mm}
r_n = \RG_n  \prod_{j= -\infty}^n \Bigl( 
\frac{1 - \QG_{j-1}\RG_{j}}{1 + \QG_{j-1}\RG_{j-1}}
\Bigr)^{2\gamma}.
\label{gtrf}
\eeq
Here $\gamma$ is a real parameter. 
It should be noticed that both $\log (1-\QG_{n-1} \RG_n)$ and 
$\log (1+\QG_n \RG_n)$ 
are conserved densities for \eref{sdGI}. 
The form of the transformation is so chosen 
that 
the reduction $\RG_n = \i \s \QRG_{n-\hf}$ results in 
$r_n = \i \s q_{n-\hf}^{\; \ast}$. 
%

The inverse of the transformation \eref{gtrf} is written as 
\beq
\QG_n = q_n \prod_{j= -\infty}^n 
\frac{g_\gamma (q_{j} r_{j}) }{g_\gamma (-q_{j-1}r_{j})} 
\hspace{8mm}
\RG_n = r_n \prod_{j= -\infty}^n 
\frac{g_\gamma (-q_{j-1} r_{j})}{g_\gamma (q_{j-1} r_{j-1})}.
\label{inv}
\eeq
Here 
$g_\gamma (y)$ is defined by the functional equation 
\[
\bigl[ 1 + y g_\gamma (y) \bigr]^{-2\gamma} = g_\gamma (y).
\]
By the successive approximation, $g_\g (y)$ 
for a general value of $\gamma$ is expressed 
as the power series:
\beq
g_\g (y) = 1 - 2\g y + \g (6\g +1) y^2 + \cdots.
\label{g_exp}
\eeq
%
Substituting \eref{inv} into 
the semi-discrete Gerdjikov--Ivanov system \eref{sdGI}, 
we obtain an integrable lattice version of the 
generalized DNLS system:
\bseq
\bea
\fl 
q_{n,t} - a q_{n+1} g_{\g}(q_{n+1}r_{n+1})
\Bigl[ \frac{1}{g_\g (-q_n r_{n+1})} - q_n r_{n+1} \Bigr]
\bigl[ 1 + (2\g+1) q_n r_n g_\g (q_n r_n) \bigr] 
\nn \\ 
\mbox{} + b q_{n-1} g_{\g}(-q_{n-1}r_{n})
\Bigl[ \frac{1}{g_\g (q_n r_{n})} + q_n r_{n} \Bigr]
\bigl[ 1 - (2\g+1) q_n r_{n+1} g_\g (-q_n r_{n+1}) \bigr] 
\nn \\
\mbox{} + (a-b) q_n -2\g b q_n r_{n+1} q_n g_\g (-q_n r_{n+1}) 
-2\g a q_n r_{n} q_n g_\g (q_n r_{n}) =0
\label{}
\eea
\bea
\fl 
r_{n,t} - b r_{n+1} g_{\g}(-q_{n}r_{n+1})
\Bigl[ \frac{1}{g_\g (q_n r_{n})} + q_n r_{n} \Bigr]
\bigl[ 1 - (2\g+1) q_{n-1} r_n g_\g (-q_{n-1} r_n) \bigr] 
\nn \\ 
\mbox{} + a r_{n-1} g_{\g}(q_{n-1}r_{n-1})
\Bigl[ \frac{1}{g_\g (-q_{n-1} r_{n})} - q_{n-1} r_{n} \Bigr]
\bigl[ 1 + (2\g+1) q_n r_{n} g_\g (q_n r_{n}) \bigr] 
\nn \\
\mbox{} + (b-a) r_n + 2\g a r_n q_{n} r_n g_\g (q_n r_{n}) 
+ 2\g b r_n q_{n-1} r_n g_\g (-q_{n-1} r_{n}) =0.
\label{}
\eea
\label{sd_Keq}
\eseq
%
%
The system \eref{sd_Keq} with $a=-b=\i$ gives 
an integrable semi-discretization of 
\eref{Keq}. 
As was expected from the construction of 
\eref{gtrf}, \eref{sd_Keq} 
with $b=a^\ast$ admits the reduction 
$r_n = \i \s q_{n-\hf}^{\; \ast}$. 
Although \eref{sd_Keq} is not an explicit expression 
for general $\gamma$, we can obtain 
approximate expressions 
by using 
the power series \eref{g_exp}. 

For some choices 
of $\g$, $g_\g (y)$ is written 
explicitly, {\it e.g.}
\bea
\eql{\fl
g_{-1} (y) = \Bigl( \frac{2}{1+ \sqrt{1-4y}}\Bigr)^2  \hspace{7mm}
g_{-\frac{1}{2}} (y) = \frac{1}{1-y} 
\hspace{7mm}
g_{-\frac{1}{4}} (y) = \sqrt{1+ \frac{1}{4} y^2} + \hf y 
\\
g_{0} (y) = 1 \hspace{7mm}
g_{\hf} (y) = \frac{2}{1+ \sqrt{1+4y}}\,.} 
\nn 
\eea
When $\gamma= -1/2$, 
\eref{sd_Keq} coincides with the semi-discrete Kaup--Newell 
system \eref{sdKN} for scalar variables. 

\subsection{Alternative semi-discrete Chen--Lee--Liu system}
\label{AlCLL}

Setting $\gamma= -1/4$ in \eref{sd_Keq}, we obtain 
an integrable 
semi-discretization of the Chen--Lee--Liu system 
which is different from \eref{sdCLL}. 
We write the equation for $\QD_n$:
%
\bea
&& \fl
\QD_{n,t} -a \bigl\{ 1 + (\QD_n \RD_n)^2 + \QD_n \RD_n 
	\sqrt{1 + (\QD_n \RD_n)^2} \bigr\}
\bigl\{ \bigl[ \sqrt{1 + (\QD_{n+1} \RD_{n+1})^2}
 + \QD_{n+1} \RD_{n+1} \bigr] 
\nn \\
&& \fl
\mbox{} \times 
 \bigl[ \sqrt{1 + (\QD_{n} \RD_{n+1})^2} - \QD_{n} \RD_{n+1} \bigr] \QD_{n+1}
 - \QD_n \bigr\}
+ b \bigl\{ 1 + (\QD_n \RD_{n+1})^2 - \QD_n \RD_{n+1} 
	\sqrt{1 + (\QD_n \RD_{n+1})^2} \bigr\}
\nn \\
&& \fl
\mbox{} \times 
	\bigl\{ \bigl[ \sqrt{1 + (\QD_{n-1} \RD_{n})^2}
 - \QD_{n-1} \RD_{n} \bigr] 
 \bigl[ \sqrt{1 + (\QD_{n} \RD_{n})^2} + \QD_{n} \RD_{n} \bigr] \QD_{n-1}
 - \QD_n \bigr\} = 0.
\label{asdCLL}
\eea
Here we have changed the scaling of $\QD_n \RD_n$ 
to eliminate fractions. 
The alternative semi-discrete Chen--Lee--Liu system 
(\eref{asdCLL} and the equation for $\RD_n$) 
with $b=a^\ast$ admits the reduction 
$\RD_n = \i \s \QRD_{n-\hf}$. 
In this respect, the alternative system 
is superior to 
the semi-discrete Chen--Lee--Liu system 
\eref{sdCLL}. 

The 
system \eref{sdCLL} for scalar variables is related to 
the semi-discrete Gerdjikov--Ivanov system \eref{sdGI} 
via the transformation 
\[
\QC_n = \QG_n \prod_{j= -\infty}^n (1 - \QG_{j-1}\RG_{j})
\hspace{8mm}
\RC_n = \RG_n  \prod_{j= -\infty}^n \frac{1}{1 - \QG_{j-1}\RG_{j}}.
\]
These relations among the semi-discrete DNLS systems 
\eref{sdKN}, \eref{sdCLL} and \eref{sdGI} 
can be generalized to the case of matrix-valued variables 
(cf \cite{TW3}). 

\section{Systems 
related to the semi-discrete DNLS}
\setcounter{equation}{0}

In this section, 
we obtain integrable 
semi-discretizations of various systems 
through transformations 
and reductions 
for the semi-discrete DNLS systems. 

\subsection{Semi-discrete mixed NLS systems}
\label{sdmNLS}

The so-called mixed NLS equations are obtained 
as a mixture of an integrable DNLS 
equation and the NLS equation \cite{WKI1,Clark,Clark2,Clark3,Sak,Kawa,Kakei}. 
%
To give an actual example, 
we add 
$+c q^2 r$ and 
$-c r^2 q$ to the 
left-hand sides of \eref{Keq}, respectively. 
Since the mixed NLS equations are connected with the 
original DNLS equations through transformations of 
variables, 
they are also integrable. 
%
%
%
By a semi-discrete version of such transformations 
%
\[
q_n \to q_n \e^{\i (\theta n  + \varphi t) }
\hspace{7mm}
r_n \to r_n \e^{-\i (\theta n  + \varphi t) }
\]
for the semi-discrete DNLS systems, 
%
we straightforwardly obtain 
integrable semi-discretizations of 
the mixed NLS systems. 
%
%
%
%

\subsection{Semi-discrete NLS equations and 
semi-discrete 
mKdV equations}
\label{SDNLS}

For special choices of the parameters $\theta, \varphi$, 
the semi-discrete mixed NLS systems are reduced to 
novel lattice versions of 
the NLS equation and the mKdV equation. 

Substituting $\QK_n = (-1)^n \e^{2(b-a)t} q_n$, 
$\RK_n = (-1)^n \e^{2(a-b)t} r_n$ into the semi-discrete 
Kaup--Newell system \eref{sdKN}, 
we obtain 
%
\bseq
\bea
&& \fl
q_{n,t} + a (I_1 - q_{n+1} r_{n+1})^{-1} q_{n+1}
+ a (I_1 - q_{n} r_{n})^{-1} q_{n}
- b (I_1 - q_{n} r_{n+1})^{-1} q_{n}
\nn \\
&& 
\mbox{}
- b (I_1 - q_{n-1} r_{n})^{-1} q_{n-1} 
+ 2(b-a) q_n = O
\\
&& \fl
r_{n,t} + b (I_2 - r_{n+1} q_{n})^{-1} r_{n+1}
+ b (I_2 - r_{n} q_{n-1})^{-1} r_{n}
- a (I_2 - r_{n} q_{n})^{-1} r_{n}
\nn \\
&& \mbox{}
- a (I_2 - r_{n-1} q_{n-1})^{-1} r_{n-1} 
+ 2 (a-b) r_n = O. 
\eea
\label{sdNLS1}
\eseq
%
%
Putting $\QC_n = (-1)^n \e^{2(b-a)t} q_n$, 
$\RC_n = (-1)^n \e^{2(a-b)t} r_n$ into the semi-discrete 
Chen--Lee--Liu system \eref{sdCLL}, we obtain
\bea
\eql{\fl
q_{n,t} + a (q_{n+1}+ q_n) (I_2 + r_{n} q_{n})
- b (q_{n} + q_{n-1}) (I_2 - r_{n} q_{n-1})^{-1} 
+ 2(b-a) q_n = O
\\
\fl
r_{n,t} + b (I_2 - r_{n+1} q_{n})^{-1} (r_{n+1} + r_{n}) 
- a (I_2 + r_{n} q_{n}) (r_{n}+ r_{n-1}) 
+ 2(a-b) r_n = O.
}
\label{sdNLS2}
\eea
%
%
%
Substitution of $\QG_n = (-1)^n \e^{2(b-a)t} q_n$, 
$\RG_n = (-1)^n \e^{2(a-b)t} r_n$ into the semi-discrete 
Gerdjikov--Ivanov system \eref{sdGI} yields 
\bseq
\bea
&& \fl
q_{n,t} + a q_{n+1} -b q_{n-1} + (b-a) q_n
+ a q_{n+1} (r_{n+1} + r_n) q_{n}
- b q_{n} (r_{n+1} + r_n) q_{n-1}
\nn \\
&& 
\mbox{}
+ a q_{n+1} r_{n+1}q_{n} r_{n} q_{n}
- b q_{n} r_{n+1}q_{n} r_{n} q_{n-1} = O
\\
&& \fl
r_{n,t} + b r_{n+1} - a r_{n-1} + (a-b) r_n
+ b r_{n+1} (q_n + q_{n-1}) r_{n}
- a r_{n} (q_n + q_{n-1}) r_{n-1}
\nn \\
&& \mbox{}
+ b r_{n+1} q_{n} r_{n} q_{n-1} r_{n}
- a r_{n} q_{n} r_{n} q_{n-1} r_{n-1} = O. 
\eea
\label{sdNLS3}
\eseq
%
%
The 
systems \eref{sdNLS1}--\eref{sdNLS3} 
give integrable semi-discretizations of the 
matrix NLS equation for $a=-b=-\i$ and 
the matrix mKdV equation for $a=b=1$, respectively. 

In the continuous theory, the matrix NLS equation \eref{NLS} and 
the matrix mKdV equation (see (4.24) in \cite{Konope} or 
(2.12) in \cite{TW1}) admit 
either of the reductions 
$v = \s u^\ast$ and $v = \s u^\dagger$. 
The system \eref{sdNLS3} with $b=a^\ast$
admits the reduction $r_n = \s q_{n-\hf}^{\; \ast}$
and not the reduction $r_n = \s q_{n-\hf}^{\; \dagger}$. 
This is 
similar to 
the property of the matrix Ablowitz--Ladik 
system \eref{sd-matrix} explained in section~\ref{GeneAL}. 
The 
system \eref{sdNLS1} with $b=a^\ast$ admits 
the reduction $r_n = \s q_{n-\hf}^{\; \dagger}$ as well as 
the reduction $r_n = \s q_{n-\hf}^{\; \ast}$. 
In this respect, 
the 
differential-difference 
scheme \eref{sdNLS1} faithfully 
inherits the property of the continuous hierarchy. 
This enables us to obtain integrable semi-discretizations of 
reductions of the matrix NLS hierarchy, {\it e.g.}\ 
the coupled NLS equations \cite{Fordy1,Manakov}, the coupled 
Hirota equations \cite{Linden2,Sak,Hir,TP} and the coupled Sasa--Satsuma 
equations (cf section~\ref{sdcss}). 

\subsection{Semi-discrete 
Sasa--Satsuma equations} 
\label{sdcss}

If 
dependent variables are real matrices, 
the Hermitian conjugation 
coincides with the transposition. 
The matrix mKdV equation in this case 
takes the following form \cite{Linden2,TW1,AF}
\beq
u_t + u_{xxx} - 3 \s (u_x {}^t \hspace{-0.5mm}u u + 
 u {}^{\, t} \hspace{-0.5mm} u u_x) = O. 
\label{rmmKdV}
\eeq
%
Let us consider the vector reduction: $u = (u_1, \ldots, u_{2m}) 
\in \mathbb{R}^{2m}$ \cite{Konope,YO}. 
Defining a set of complex variables by 
$\psi_k = u_{2k-1} + \i u_{2k}$ $(k=1, \ldots, m)$, we obtain 
the coupled Sasa--Satsuma 
equations \cite{YO,Sak,SS,NPSM} (cf (4.26) in \cite{Konope}), 
%
\beq
\vecvar{\psi}_t + \vecvar{\psi}_{xxx} 
- 3 \s \| \vecvar{\psi} \|^2 \vecvar{\psi}_x 
- \frac{3}{2} \s (\| \vecvar{\psi} \|^2)_x \vecvar{\psi} = \vecvar{0}.
\label{cSS}
\eeq
Here $\vecvar{\psi} = (\psi_1, \ldots, \psi_{m}) \in \mathbb{C}^{m}$. 

The system \eref{sdNLS1} with $a=b=1$ 
and $r_n = \sigma {}^{\, t} \hspace{-0.5mm}
q_{n-\hf}$ gives an integrable semi-discretization of 
the real matrix mKdV equation \eref{rmmKdV}. 
Following the same procedure as in the continuous case, we 
obtain an integrable lattice version of the coupled 
Sasa--Satsuma equations \eref{cSS}. 

\subsection{Semi-discrete KdV equations}
\label{SDKDV}

If we consider the reduction $a=b=1$, $q_n = I +p_n$, $r_n=I$ for 
\eref{sdNLS1}, we obtain a matrix generalization of a 
semi-discrete KdV equation \cite{Wad,Hiro},
\[
p_{n,t} = p_{n+1}^{\, -1} - p_{n-1}^{\, -1}.
\]
Through the reduction $a=b=1$, 
$q_n = -I + y_n$, $r_n=I$ for 
\eref{sdNLS3}, we obtain a matrix version of a 
semi-discrete KdV equation \cite{Wad},
\[
y_{n,t} + y_{n+1} y_n^{\, 2} - y_n^{\, 2} y_{n-1} =O.
\]

\subsection{Semi-discrete Burgers systems}

The matrix Burgers equation,
\beq
\QC_t - \QC_{xx} + 2\QC_x \QC = O
\label{Beq}
\eeq
is obtained as the reduction $\RC = -I$ for the 
Chen--Lee--Liu equation \eref{CLLeq2} (after scalings). 
The Hopf--Cole transformation $\QC = -F_x F^{-1}$ casts 
the linear diffusion equation $F_t - F_{xx} =O$ into \eref{Beq}.

In a similar way, we obtain integrable semi-discretizations 
of the Burgers equation 
together with its higher symmetry. 
Equating $\RC_n$ with $-I$ in the semi-discrete 
Chen--Lee--Liu system \eref{sdCLL}, we obtain 
\beq
\QC_{n,t} - a(\QC_{n+1} - \QC_{n})(I - \QC_{n})
- b (\QC_{n} - \QC_{n-1})
(I - \QC_{n-1})^{-1} = O.
\label{sdB1}
\eeq
The 
substitution 
$
I - \QC_{n} = F_{n+1} F_n^{-1}
$
relates \eref{sdB1} with 
the linear equation
\[
F_{n,t} - a F_{n+1} + b F_{n-1} + (a-b) F_n =O.
\]
Setting $\RD_n =-1$ in 
the alternative semi-discrete Chen--Lee--Liu system \eref{asdCLL}, we obtain 
%
\bea
&& \fl
\QD_{n,t} - \sqrt{1 + \QD_n^{\, 2}}
\bigl\{ a \bigl[ \sqrt{1 + \QD_{n+1}^{\; 2}} 
	- \QD_{n+1} \bigr] \QD_{n+1}
 - a\bigl[ \sqrt{1 + \QD_{n}^{\, 2}} - \QD_{n} \bigr] \QD_n 
\nn \\
&& \mbox{} 
 + b \bigl[ \sqrt{1 + \QD_{n}^{\, 2}} + \QD_{n} \bigr] \QD_n 
 - b \bigl[ \sqrt{1 + \QD_{n-1}^{\; 2}} + \QD_{n-1} \bigr] \QD_{n-1}
\bigr\}= 0.
\label{sdB2}
\eea
The substitution 
$[ \sqrt{1 + \QD_n^{\, 2}} - \QD_n ]^2 = f_{n+1} f_n^{-1}$ 
connects \eref{sdB2} with the linear equation
\[
f_{n,t} - a f_{n+1} + b f_{n-1} + (a-b) f_n =0.
\]
The 
systems \eref{sdB1} and \eref{sdB2} give 
two lattice versions of the Burgers equation 
for $a=-b=1$ 
and its higher symmetry for $a=b=1$, respectively. 

\section{Semi-discrete massive Thirring-type models} 
\setcounter{equation}{0}
\label{sdMTM}

In section~\ref{sdDNLS}, we obtained the Lax pairs 
for the semi-discrete DNLS systems. 
If we can replace the $M_n$-matrices with appropriate ones, 
we obtain integrable 
semi-discretizations of the massive Thirring-type models. 
%
For this purpose, the Lax pair introduced in~\ref{appB} 
provides useful information. 
%

\subsection{Semi-discrete Kaup--Newell type}
\label{sdKNm}

For the semi-discrete Kaup--Newell system in section~\ref{SubKN}, 
we replace the $M_n$-matrix \eref{Mn_KN} 
with the following one:
\beq
\MK_n = \frac{1}{z - \frac{1}{z}} \left[
\begin{array}{cc}
 \frac{1}{z} I_1 -2 (z-\frac{1}{z}) \PHK_n \CHK_n & 2\i f(z) \PHK_n  \\
\frac{2 \i}{f(z)} (-1 + \frac{1}{z^2}) \CHK_n 
	& -z I_2 + 2(z-\frac{1}{z}) \CHK_n \PHK_n \\
\end{array}
\right].
\label{KNmass}
\eeq
Here $\PHK_n$ and $\CHK_n$ are new variables 
which are, respectively, $l_1 \times l_2$ and 
$l_2 \times l_1$ matrices. 
Substituting the Lax pair \eref{Ln_KN} and \eref{KNmass} 
into the zero-curvature condition \eref{Lax_eq}, we obtain 
\beq
\eql{
\QK_{n,t} + \i (\PHK_n+ \PHK_{n+1}) 
	+ 2 (\QK_n \CHK_{n+1} \PHK_n + \PHK_{n+1} \CHK_{n+1} \QK_n) = O
\\
\RK_{n,t} - \i (\CHK_n+ \CHK_{n+1}) 
	- 2 (\RK_n \PHK_{n} \CHK_n + \CHK_{n+1} \PHK_{n} \RK_n) = O
\\
\PHK_{n} - \PHK_{n+1} + \i \QK_n =O
\\
\CHK_{n} - \CHK_{n+1} - \i \RK_n =O.
}
\label{KNm}
\eeq
The system \eref{KNm} gives an integrable semi-discretization 
of the massive Thirring model of the Kaup--Newell type 
(see (A.4) in \cite{TW3}). As we expect from the property 
of the semi-discrete Kaup--Newell system \eref{sdKN}, 
\eref{KNm} admits either the reduction 
$\RK_n = \i \s \QRK_{n-\hf}$, $\CHK_n = \i \s \PCK_{n-\hf}$
or the reduction 
$\RK_n = \i \s \QHK_{n-\hf}$, $\CHK_n = \i \s \PMK_{n-\hf}$. 

We can eliminate $\QK_n$ and $\RK_n$ 
to obtain an integrable semi-discretization of the Mikhailov 
model (see (4.21) in \cite{GIK} or (3.47) in \cite{Linden1}),
\beq
\eql{
\PHK_{n,t} - \PHK_{n+1,t} + \PHK_n + \PHK_{n+1} 
	+ 2 (\PHK_n \CHK_{n+1} \PHK_n - \PHK_{n+1} \CHK_{n+1} \PHK_{n+1}) =O
\\
\CHK_{n,t} - \CHK_{n+1,t} + \CHK_n + \CHK_{n+1} 
	- 2 (\CHK_n \PHK_{n} \CHK_n - \CHK_{n+1} \PHK_{n} \CHK_{n+1}) =O.
}
\label{sdMik}
\eeq
Through the reduction $\PHK_n = (-1)^n (\omega_n + I/4)$, 
$\CHK_n = (-1)^n I$, \eref{sdMik} collapses into the 
simplest version of the dressing chain \cite{SY},
\[
\omega_{n+1,t} + \omega_{n,t} +2 \omega_{n+1}^{\, 2} - 2\omega_{n}^{\, 2}=O.
\]

\subsection{Semi-discrete Chen--Lee--Liu type}
For the semi-discrete Chen--Lee--Liu system in section~\ref{SubCLL}, 
we replace the $M_n$-matrix \eref{Mn_CLL} 
with the following one:
\beq
\MC_n = \frac{1}{z - \frac{1}{z}} \left[
\begin{array}{cc}
 \frac{1}{z} I_1 & 2\i f(z) \PHC_n \\
\frac{2 \i}{f(z)} (-1 + \frac{1}{z^2}) \CHC_n 
	& -\frac{1}{z} I_2 + 2(z-\frac{1}{z}) \CHC_n \PHC_n \\
\end{array}
\right].
\label{CLLmass}
\eeq
Plunging the Lax pair \eref{Ln_CLL} and \eref{CLLmass} into 
the zero-curvature condition \eref{Lax_eq}, we obtain
\beq
\eql{
\QC_{n,t} + 2 \i \PHC_n + 2 \QC_n \CHC_n \PHC_n = O
\\
\RC_{n,t} - 2 \i \CHC_{n+1} - 2 \CHC_{n+1} \PHC_{n+1} \RC_n = O
\\
\PHC_{n} - \PHC_{n+1} + \i \QC_n + \QC_n \RC_n \PHC_n =O
\\
\CHC_{n} - \CHC_{n+1} - \i \RC_n - \CHC_{n+1} \QC_{n} \RC_n =O.
}
\label{CLLm}
\eeq
The system \eref{CLLm} gives an integrable semi-discretization 
of the massive Thirring model 
(see (A.3) in \cite{TW3}). 
As we expect from the property 
of the semi-discrete Chen--Lee--Liu system \eref{sdCLL}, 
\eref{CLLm} admits neither the reduction 
$\RC_n = \i \s \QRC_{n-k}$, $\CHC_n = \i \s \PCC_{n-k}$
nor the reduction 
$\RC_n = \i \s \QHC_{n-k}$, $\CHC_n = \i \s \PMC_{n-k}$. 
In section~\ref{AsdCLL}, we obtain 
an alternative semi-discretization of 
the massive Thirring model without this drawback.  

\subsection{Semi-discrete Gerdjikov--Ivanov type}

For the semi-discrete Gerdjikov--Ivanov system 
in section~\ref{SubGI}, 
we replace the $M_n$-matrix \eref{Mn_GI} 
with the following one:
\beq
\fl
\MG_n = \frac{1}{z - \frac{1}{z}} \left[
\begin{array}{cc}
 \frac{1}{z} I_1 & 2\i f(z) \PHG_n  \\
\frac{2 \i}{f(z)} (-1 + \frac{1}{z^2}) (I_2 - \RG_n \QG_{n-1})\CHG_n 
	& -z I_2 + 2\i(z-\frac{1}{z}) \RG_n \PHG_n \\
\end{array}
\right].
\label{GImass}
\eeq
Putting the Lax pair \eref{LG_n} and \eref{GImass} into 
the zero-curvature condition \eref{Lax_eq}, we obtain
\beq
\eql{
\QG_{n,t} + \i (\PHG_n + \PHG_{n+1}) + 
	\i (\QG_n \RG_n \PHG_n - \PHG_{n+1} \RG_{n+1} \QG_n)= O
\\
\RG_{n,t} - \i (\CHG_n + \CHG_{n+1}) +
	\i (\RG_n \QG_{n-1} \CHG_n - \CHG_{n+1} \QG_{n} \RG_n)= O
\\
\PHG_{n} - \PHG_{n+1} + \i \QG_n + (\QG_n \RG_n \PHG_n 
	+ \PHG_{n+1} \RG_{n+1} \QG_n)=O
\\
\CHG_{n} - \CHG_{n+1} - \i \RG_n - (\RG_n \QG_{n-1} \CHG_n 
	+ \CHG_{n+1} \QG_{n} \RG_n)=O.
}
\label{GIm}
\eeq
The system \eref{GIm} gives an integrable semi-discretization 
of the massive Thirring model of the Gerdjikov--Ivanov type 
(see (A.4) in \cite{TW3} with interchanging $t$ and $x$). 
As we expect from the property of the semi-discrete 
Gerdjikov--Ivanov system \eref{sdGI}, \eref{GIm} admits the reduction 
$\RG_n = \i \s \QRG_{n-\hf}$, $\CHG_n = \i \s \PCG_{n-\hf}$
and not the reduction 
$\RG_n = \i \s \QHG_{n-\hf}$, $\CHG_n = \i \s \PMG_{n-\hf}$. 

If we eliminate $\PHG_n$ and $\CHG_n$, we obtain 
another semi-discretization of the Mikhailov model,
%
\beq
\eql{
\QG_{n+1,t} (I_2 - \RG_{n+1} \QG_n) - (I_1 + \QG_{n+1} \RG_{n+1}) 
	\QG_{n,t} - \QG_n - \QG_{n+1} =O
\\
\RG_{n+1,t} (I_1 + \QG_{n} \RG_n) - (I_2 - \RG_{n+1} \QG_{n}) 
	\RG_{n,t} - \RG_n - \RG_{n+1} =O.
}
\label{sdMik2}
\eeq
Comparing \eref{sdMik2} with \eref{sdMik}, we notice 
the interchange of space and time. 
Through the reduction $\QG_n=(-1)^n (\xi_n - I)$, $\RG_n = (-1)^n I$ 
for \eref{sdMik2}, we obtain the following equation:
\[
(\xi_{n+1} \xi_n)_t = \xi_{n+1} - \xi_n.
\]

\subsection{Alternative semi-discrete Chen--Lee--Liu type}
\label{AsdCLL}

We assume that dependent variables are scalar and 
consider the transformation \eref{gtrf} together with 
%
\beq
\fl
\phi_n = \PHG_n \prod_{j= -\infty}^n \Bigl( 
\frac{1 - \QG_{j-1}\RG_{j}}{1 + \QG_{j-1}\RG_{j-1}}
\Bigr)^{-2\gamma} \hspace{7mm}
\chi_n = \CHG_n  \prod_{j= -\infty}^n \Bigl( 
\frac{1 - \QG_{j-2}\RG_{j-1}}{1 + \QG_{j-1}\RG_{j-1}}
\Bigr)^{2\gamma}.
\label{gtrf2}
\eeq
%
Substituting the inverse transformation into 
\eref{GIm}, we obtain an integrable 
semi-discretization of a generalized massive Thirring 
model. 
When $\g=-1/2$, the obtained 
system coincides with \eref{KNm}. 
Setting $\g = -1/4$, we obtain an 
integrable semi-discretization of 
the massive Thirring model which is different from \eref{CLLm}. 
We write the equations for $q_n$ and $\PHD_n$: 
\bseq
\bea
\fl
q_{n,t} + \i \bigl[ \sqrt{ 1 + (q_n r_n)^2} + q_n r_n\bigr] \PHD_n 
	+ \i \bigl[ \sqrt{ 1 + (q_n r_{n+1})^2} - q_n r_{n+1}\bigr]\PHD_{n+1} 
\nn \\
\fl	\mbox{}+ 2 \bigl[ \sqrt{ 1 + (q_n r_{n})^2} + q_n r_{n} \bigr] 
	\PHD_{n} \CHD_{n+1} q_n + 2 \bigl[ \sqrt{ 1 + (q_n r_{n+1})^2} 
	- q_n r_{n+1} \bigr] \PHD_{n+1} \CHD_{n+1} q_n =0
\nn \\
\eea
\beq
\fl
\bigl[ \sqrt{ 1 + (q_n r_n)^2} + q_n r_n\bigr] \PHD_n  
	- \bigl[ \sqrt{ 1 + (q_n r_{n+1})^2} -q_n r_{n+1} \bigr] \PHD_{n+1} 
	+ \i q_n =0.
\eeq
\label{sdmT}
\eseq
Here we have changed the scaling of $q_n r_n$ to eliminate fractions. 
Unlike \eref{CLLm}, the alternative semi-discrete massive Thirring model 
(\eref{sdmT} and the equations for $\RD_n$ and $\CHD_n$) 
admits the reduction $\RD_n = \i \s 
\QRD_{n-\hf}$, $\CHD_n = \i \s \PCD_{n-\hf}$.

\section{Concluding remarks}
\label{}
\setcounter{equation}{0}

In this paper, we have 
investigated 
integrable discretizations of the DNLS systems 
from the point of view of a Lax-pair formulation. 
We have found that lattice versions of the Kaup--Newell 
system, the Chen--Lee--Liu system 
and the Gerdjikov--Ivanov system are embedded 
in a generalization of the Ablowitz--Ladik system. 
With appropriate gauge transformations, the 
eigenvalue problems 
for the lattice systems 
coincide with the continuous counterparts in the continuum limit. 
Using a discrete analogue of the phase transformation \eref{cgt}, 
we obtain a lattice version 
of the generalized DNLS system by Kund. 
%
All the discrete DNLS systems but the semi-discrete 
Chen--Lee--Liu system \eref{sdCLL} 
admit the reduction of complex conjugation 
between two dependent variables. 
%
This property is indispensable for 
applications  
such as difference schemes for numerical computation or 
modelling of 
nonlinear lattice vibrations. 
We stress that the discrete DNLS systems 
have relation to a variety of discrete integrable systems. 
Through changes of variables 
and reductions, 
we obtain 
integrable discretizations of 
the 
mixed NLS, matrix NLS, matrix KdV, matrix mKdV, 
coupled Sasa--Satsuma, Burgers equations, etc. 
From the eigenvalue problems for the discrete DNLS systems, 
we derive integrable discretizations of the massive Thirring-type models.  

Besides the Ablowitz--Ladik system, 
there are other integrable discretizations of the 
NLS hierarchy. 
We conjecture that other 
discrete DNLS systems are embedded in the discrete NLS systems. 
In fact, 
we obtain another semi-discretization of the 
Gerdjikov--Ivanov system 
in connection with the semi-discrete NLS system 
studied in \cite{Merola} (see also \cite{Date,GI2}). 
However, the obtained system 
does not admit the reduction of complex conjugation. 
Thus it is less attractive than \eref{sdGI}. 
The example implies that the Ablowitz--Ladik formulation 
is an optimal starting point for the study of discrete DNLS systems. 

%
The NLS and DNLS systems 
for scalar variables 
possess tri-Hamiltonian structure \cite{Ferr,Fan}. 
The three sets of 
Poisson brackets 
are, respectively, 
the $\de{'}(x-y)$ type for the Kaup--Newell system, 
the $\de(x-y)$ type for the Chen--Lee--Liu system 
and the $\de(x-y)$ type for the NLS system. 
We should note that all these systems are connected 
via transformations of the dependent variables. 
We infer that 
the lattice versions of these systems studied in this paper 
possess tri-Hamiltonian structure. 
%
The discrete counterpart of the first 
Hamiltonian structure is 
\eref{firstH} for the semi-discrete Kaup--Newell 
system \eref{sdKN}. 
The Ablowitz--Ladik system \eref{sd-matrix} 
has a set of ultra-local Poisson brackets \cite{Kako,Kulish,Suris} which is 
the discrete analogue of the third 
Hamiltonian structure. 
Since the lattice systems 
are each connected with the others through changes of 
variables (see section~\ref{sdDNLS}), 
they are at least bi-Hamiltonian. 
We have not found the discrete counterpart of the second 
Hamiltonian structure for either the semi-discrete 
Chen--Lee--Liu system \eref{sdCLL} 
or the alternative 
system \eref{asdCLL}.
It remains an unsolved problem to prove explicitly that 
the Ablowitz--Ladik system and the semi-discrete 
DNLS systems are tri-Hamiltonian. 

\ack
The author 
would like to thank 
Prof R Sasaki and 
Dr Y Fujii for helpful suggestions in improving the manuscript. 
This research was supported in part by 
a JSPS 
Fellowship 
for Young Scientists.

\appendix

\section{Semi-discrete sine-Gordon equation 
as a negative flow of the Ablowitz--Ladik hierarchy}
\label{appB}

In this appendix, we prove that a 
semi-discrete 
sine-Gordon equation 
and its generalization 
are integrable via the ISM 
based on the Ablowitz--Ladik eigenvalue problem. 
%
%
We introduce the following form of the Lax pair:
\beq
L_n = \left[
\begin{array}{cc}
 z & u_n \\
 v_n & \frac{1}{z}\\
\end{array}
\right]
\hspace{7mm}
M_n = \frac{1}{z - \frac{1}{z}}
\left[
\begin{array}{cc}
\frac{1}{z} a_n & b_n \\
 c_n & -z a_n \\
\end{array}
\right]. 
\label{sGM}
\eeq
Here $z$ is the spectral parameter. $u_n$, $v_n$, $a_n$, $b_n$, $c_n$ 
are scalar variables. 
The $M_n$-matrix is simpler than 
the one by Ablowitz and Ladik \cite{AL1}, 
while the $L_n$-matrix is the same. 
%
Substituting \eref{sGM} into the zero-curvature 
condition \eref{Lax_eq}, we obtain the following six equations:
\bseq
\bea
a_n - a_{n+1} + u_n c_n - v_n b_{n+1} =0
\label{an}
\\
u_n (c_n - c_{n+1}) + v_n (b_n - b_{n+1}) =0
\label{uvn}
\\
b_n - b_{n+1} - u_n (a_n + a_{n+1}) =0
\label{bn}
\\
c_n - c_{n+1} + v_n (a_n + a_{n+1}) =0
\label{cn}
\\
u_{n,t} + b_n - u_n a_n =0
\label{un}
\\
v_{n,t} - c_n - v_n a_n = 0.
\label{vn}
\eea
\eseq
Thanks to \eref{bn} and \eref{cn}, we can write $u_n$ and 
$v_n$ in terms of $a_n$, $b_n$, $c_n$:
\beq
u_n = \frac{b_n - b_{n+1}}{a_n + a_{n+1}} 
\hspace{7mm}
v_n = -\frac{c_n - c_{n+1}}{a_n + a_{n+1}}.
\label{uvabc}
\eeq
Then \eref{uvn} is automatically satisfied. 
Putting \eref{uvabc} into \eref{an}, we obtain 
\beq
a_n^{\, 2} + b_n c_n = \eta^2
\label{abceta}
\eeq
where $\eta$ is a constant. 

If we set 
\[
a_n = \eta \cos \theta_n \hspace{7mm} 
b_n = c_n = \eta \sin \theta_n  
\]
$u_n$ and $v_n$ 
are expressed as
\[
u_n = -v_n = \frac{\sin \theta_n - \sin \theta_{n+1}} 
{\cos \theta_n + \cos \theta_{n+1}}=
\tan \Bigl( \frac{\theta_n - \theta_{n+1}}{2} \Bigr).
\]
Then \eref{un} and \eref{vn} are combined into 
a semi-discrete sine-Gordon equation \cite{Levi,Pilloni},
\beq
\theta_{n+1,t} - \theta_{n,t} = \eta (\sin \theta_{n} + \sin \theta_{n+1}).
\label{sdSG1}
\eeq
By defining $\a_n$ by $\theta_n = (\a_{n} + \a_{n+1})/2$, 
we rewrite \eref{sdSG1} as \cite{Orf}
\beq
\a_{n+1,t} - \a_{n,t} = 2 \eta \sin \Bigl(\frac{\a_{n}+ \a_{n+1}}{2} \Bigr)
\label{sdSG2}
\eeq
under appropriate boundary conditions. 

More generally, $b_n$ and $c_n$ 
are functionally independent. 
Substituting $a_n = \sqrt{\eta^2 - b_n c_n}$ 
into \eref{uvabc}, \eref{un} and \eref{vn}, we obtain a 
semi-discretization of the reduced equation for the 
O(4) nonlinear $\s$-model \cite{Eich},
\bseq
\bea
\fl
\left(
\frac{b_n - b_{n+1}}{\sqrt{\eta^2 - b_n c_n} 
+ \sqrt{\eta^2 - b_{n+1} c_{n+1}}}
\right)_t + 
\frac{b_n \sqrt{\eta^2 - b_{n+1} c_{n+1}} + b_{n+1}\sqrt{\eta^2 - b_n c_n}}
{\sqrt{\eta^2 - b_n c_n} + \sqrt{\eta^2 - b_{n+1} c_{n+1}}} =0
\\
\fl
\left(
\frac{c_n - c_{n+1}}{\sqrt{\eta^2 - b_n c_n} 
+ \sqrt{\eta^2 - b_{n+1} c_{n+1}}}
\right)_t + 
\frac{c_n \sqrt{\eta^2 - b_{n+1} c_{n+1}} + c_{n+1}\sqrt{\eta^2 - b_n c_n}}
{\sqrt{\eta^2 - b_n c_n} + \sqrt{\eta^2 - b_{n+1} c_{n+1}}} =0.
\eea
\label{sdsig}
\eseq
If we set $\eta=0$ and $b_n= -c_n=\e^{\phi_n}$, 
\eref{sdsig} is simplified to a semi-discrete Liouville equation \cite{Hirot},
\[
\phi_{n+1,t} - \phi_{n,t} = \e^{\phi_n} + \e^{\phi_{n+1}}.
\]

We have shown that 
the semi-discrete sine-Gordon equation \eref{sdSG1} and 
its generalization \eref{sdsig} are associated with 
the Ablowitz--Ladik eigenvalue problem. 
Thus they 
are integrated by the ISM straightforwardly. 
The semi-discrete sine-Gordon equation \eref{sdSG1} 
is associated with 
another ISM-solvable 
eigenvalue problem \cite{Levi,Pilloni}. 
It is now clear that the difference 
between the two eigenvalue problems is not 
essential (see also \cite{Ishimori2}). 
We obtain a lot of knowledge 
on the semi-discrete sine-Gordon equation and its generalization 
from the study of the Ablowitz--Ladik 
hierarchy. 

In section~\ref{sdMTM}, we investigate integrable semi-discretizations of 
the massive Thirring-type models. 
The $M_n$-matrix in \eref{sGM} 
with the constraint \eref{abceta}
gives helpful information for that. 

\section{Time discretizations of semi-discrete sine-Gordon equation} 
\label{appC}

In this appendix, we discuss time discretizations of the 
semi-discrete sine-Gordon equation and its generalization 
obtained in~\ref{appB}. 
We consider a time discretization of \eref{line}:
\beq
\Psi_{n+1} = L_n \Psi_n \hspace{7mm} \ti{\Psi}_n = V_n \Psi_n
\label{line2}
\eeq
where the tilde 
denotes the step-up shift ($l \to l+1$) 
in the discrete time $l \in {\mathbb Z}$. 
The compatibility of \eref{line2} leads to 
a full-discrete version of the zero-curvature 
condition \cite{Suris,AL3}:
\beq
\ti{L}_n V_n = V_{n+1} L_n.
\label{Lax_full}
\eeq
As a discrete-time analogue of (\ref{sGM}), 
we introduce the following form of the Lax pair: 
\beq
L_n = \left[
\begin{array}{cc}
 z & u_n \\
 v_n & \frac{1}{z}\\
\end{array}
\right]
\hspace{7mm}
V_n = 
\left[
\begin{array}{cc}
1 & 0\\
0 & 1\\
\end{array}
\right] + 
\frac{h}{z - \frac{1}{z}}
\left[
\begin{array}{cc}
\frac{1}{z} a_n & b_n \\
 c_n & -z a_n \\
\end{array}
\right].
\label{fsGV}
\eeq
Here $h$ denotes the difference interval of time. 
Substituting \eref{fsGV} into the zero-curvature 
condition \eref{Lax_full}, we obtain 
the following six equations:
\bseq
\bea
a_n - a_{n+1} + \ti{u}_n c_n - v_n b_{n+1} =0
\label{fan}
\\
\ti{u}_n c_n - u_n c_{n+1} + \ti{v}_n b_n - v_n b_{n+1} =0
\label{fuvn}
\\
b_n - b_{n+1} - \ti{u}_n a_n - u_n a_{n+1} =0
\label{fbn}
\\
c_n - c_{n+1} + \ti{v}_n a_n + v_n a_{n+1} =0
\label{fcn}
\\
\ti{u}_{n} - u_n + h(b_n - \ti{u}_n a_n) =0
\label{fun}
\\
\ti{v}_{n} - v_n - h(c_n + \ti{v}_n a_n) = 0.
\label{fvn}
\eea
\label{zenbu}
\eseq
Thanks to \eref{fbn} and \eref{fun}, we can write $u_n$ and $\ti{u}_n$ in 
terms of $a_n$ and $b_n$:
\bseq
\beq
u_n = \frac{b_n - b_{n+1} + h a_n b_{n+1}}{a_n + a_{n+1} -h a_n a_{n+1}}
\hspace{7mm}
\ti{u}_n = \frac{b_n - b_{n+1} - h a_{n+1} b_{n}}{a_n + a_{n+1} 
  -h a_n a_{n+1}}.
\label{utiu}
\eeq
Similarly, using \eref{fcn} and \eref{fvn}, 
we express $v_n$ and $\ti{v}_n$ in terms of $a_n$ and $c_n$:
\beq
v_n = - \frac{c_n - c_{n+1} + h a_n c_{n+1}}{a_n + a_{n+1} -h a_n a_{n+1}}
\hspace{7mm}
\ti{v}_n = -\frac{c_n - c_{n+1} - h a_{n+1} c_{n}}
	{a_n + a_{n+1} -h a_n a_{n+1}}.
\label{vtiv}
\eeq
\label{uvti}
\eseq
It is easily seen that \eref{fuvn} is automatically 
satisfied. Putting \eref{uvti} into 
\eref{fan}, we obtain 
$a_n^{\, 2} + b_n c_n = \eta^2 (1-h a_n)$ 
or, equivalently
\beq
\bigl( a_n+ \frac{h \eta^2}{2} \bigr)^{\, 2} + b_n c_n 
= \eta^2 \bigl[ 1 + \bigl(\frac{h \eta}{2} \bigr)^2 \bigr].
\label{abc_re}
\eeq
Here $\eta$ is a constant. 

The relations \eref{uvti} lead to the following system, 
\bseq
\beq 
\frac{\ti{b}_n - \ti{b}_{n+1} + h \ti{a}_n \ti{b}_{n+1}}
{\ti{a}_n + \ti{a}_{n+1} -h \ti{a}_n \ti{a}_{n+1}}
- \frac{b_n - b_{n+1} - h a_{n+1} b_{n}}{a_n + a_{n+1} -h a_n a_{n+1}} =0
\label{fdsi1}
\eeq
\beq 
\frac{\ti{c}_n - \ti{c}_{n+1} + h \ti{a}_n \ti{c}_{n+1}}
{\ti{a}_n + \ti{a}_{n+1} -h \ti{a}_n \ti{a}_{n+1}}
- \frac{c_n - c_{n+1} - h a_{n+1} c_{n}}{a_n + a_{n+1} -h a_n a_{n+1}} =0
\label{fdsi2}
\eeq
\label{fdsi}
\eseq
where $a_n$, $b_n$, $c_n$ are related through \eref{abc_re}. 
Substituting 
\[
a_n = 
-\frac{h \eta^2}{2}
  + \eta\sqrt{1 + \bigl(\frac{h \eta}{2} \bigr)^{2}} \cos \theta_n 
	\hspace{7mm} b_n = c_n = \eta \sqrt{1 + 
	\bigl(\frac{h \eta}{2} \bigr)^{2}} \sin \theta_n 
\]
into \eref{fdsi}, 
we obtain an integrable time discretization of 
the semi-discrete sine-Gordon equation \eref{sdSG1}. 
More generally, putting 
\[
a_n = 
-\frac{h \eta^2}{2}
  + \sqrt{\eta^2 + \bigl(\frac{h \eta^2}{2} \bigr)^{2} - b_n c_n} 
\]
into \eref{fdsi}, we obtain an integrable time discretization of the 
semi-discrete 
reduced nonlinear $\s$-model \eref{sdsig}. 

We shall extract a celebrated 
full-discrete sine-Gordon 
equation from \eref{zenbu}. The equation is interpreted as 
an integrable time discretization of \eref{sdSG2}. 
We introduce 
the following parametrization:
\bseq
\bea
a_n = 
-\frac{h \eta^2}{2}
  + \eta\sqrt{1 + \bigl(\frac{h \eta}{2} \bigr)^{2}} \cos 
	\Bigl( \frac{\ti{\a}_n + \a_{n+1}}{2} \Bigr)
\\
b_n = c_n = \eta \sqrt{1 + 
	\bigl(\frac{h \eta}{2} \bigr)^{2}} 
	\sin \Bigl( \frac{\ti{\a}_n + \a_{n+1}}{2} \Bigr)
\\
u_n = -v_n = 
\tan \Bigl( \frac{\a_n - \a_{n+2}}{4} \Bigr).
\eea
\label{abcuv}
\eseq
%
It is sufficient 
to consider 
\eref{fan}, \eref{fbn}, \eref{fun} among \eref{zenbu}. 
For brevity, we employ the following abbreviations: 
\[
x_n = \frac{\ti{\a}_n + \a_{n+1}}{4}
\hspace{7mm}
y_n = \frac{\a_n + \ti{\a}_{n+1}}{4}
\hspace{7mm}
k = 1+ \frac{(h \eta)^2}{2} + h \eta \sqrt{1+ \bigl( 
	\frac{h \eta}{2} \bigr)^2}.
\]
Under the parametrization \eref{abcuv}, we rewrite \eref{fan}, 
\eref{fbn}, \eref{fun} respectively as
\bseq
\beq
\frac{\cos x_n \cos y_{n+1}}{\cos(x_n - y_{n+1})} 
 = \frac{\cos y_n \cos x_{n+1}}{\cos(y_n -x_{n+1})}
\label{ccc1}
\eeq
\beq
\fl
\frac{k \sin x_n \cos y_{n+1} +  \cos x_n \sin y_{n+1}}{\cos(x_n - y_{n+1})} 
 = \frac{\sin y_n \cos x_{n+1} + k \cos y_n \sin x_{n+1}}{\cos(y_n -x_{n+1})}
\label{ccc2}
\eeq
\beq
\fl
\frac{k \sin x_n \cos y_{n+1} -k^{-1} \cos x_n \sin y_{n+1}}
{\cos(x_n - y_{n+1})} 
 = \frac{\sin y_n \cos x_{n+1} - \cos y_n \sin x_{n+1}}{\cos(y_n -x_{n+1})}.
\label{ccc3}
\eeq
\label{ccc}
\eseq
With the help of \eref{ccc1}, 
we combine \eref{ccc2} and \eref{ccc3} into one equation:
\beq
k \tan x_n = \tan y_n.
\label{tantan}
\eeq
Conversely, 
all of the relations \eref{ccc} hold 
if $x_n$ and $y_n$ satisfy \eref{tantan}. 
Equation \eref{tantan} is 
the celebrated full-discrete sine-Gordon equation \cite{Scho}. 
It is easily cast into the standard form \cite{Hirotas},
\[
\fl
\sin \Bigl( \frac{\a_n + \ti{\a}_{n+1} - \ti{\a}_n - \a_{n+1}}{4}
\Bigr) = \frac{\frac{h \eta}{2}}{\sqrt{1+ (\frac{h\eta}{2} )^2 }}
\sin \Bigl( \frac{\a_n + \ti{\a}_{n+1} + \ti{\a}_n + \a_{n+1}}{4} \Bigr).
\]

We have obtained integrable full-discretizations of the 
sine-Gordon equation and the reduced nonlinear $\s$-model. 
The obtained equations are integrable via the ISM based on 
the Ablowitz--Ladik eigenvalue problem. 

\end{document}